\documentclass[acmsmall,natbib,screen=true,nonacm]{acmart}

\usepackage[inline]{enumitem}
\usepackage{booktabs} %
\usepackage{algorithm, algpseudocode}
\usepackage{amsmath}
\usepackage{graphics}
\usepackage{epsfig}
\usepackage{graphicx}
\usepackage{xcolor}
\usepackage[skip=0pt]{caption}
\usepackage{subcaption}
\usepackage{balance}
\usepackage{multirow}
\usepackage{mathrsfs}
\usepackage{acronym}
\usepackage{placeins}
\usepackage{xcolor}
\usepackage{enumitem}
\usepackage{fancyhdr}
\usepackage{bbm}
\usepackage{tikz}
\usepackage{longtable}
\usepackage{lscape}
\usepackage{threeparttable}

\AtBeginDocument{%
  \providecommand\BibTeX{{%
    \normalfont B\kern-0.5em{\scshape i\kern-0.25em b}\kern-0.8em\TeX}}}

\newcommand{\headernodot}[1]{\vspace*{1mm}\noindent\textbf{#1}}
\newcommand{\header}[1]{\headernodot{#1.}}

\makeatletter
\def\authornotetext#1{
\if@ACM@anonymous\else
    \g@addto@macro\@authornotes{
    \stepcounter{footnote}\footnotetext{#1}}
\fi}
\makeatother

\author{Shiguang Wu}
\authornote{Both authors contributed equally to this research.}
\affiliation{
  \institution{Shandong University}
  \city{Qingdao}
  \country{China}
}
\email{shiguang.wu@mail.sdu.edu.cn}

\author{Xin Xin}
\authornotemark[1]
\affiliation{
  \institution{Shandong University}
  \city{Qingdao}
  \country{China}
}
\email{xinxin@sdu.edu.cn}

\author{Pengjie Ren}
\affiliation{
  \institution{Shandong University}
  \city{Qingdao}
  \country{China}
}
\email{jay.ren@outlook.com}

\author{Zhumin Chen}
\affiliation{
  \institution{Shandong University}
  \city{Qingdao}
  \country{China}
}
\email{chenzhumin@sdu.edu.cn}

\author{Jun Ma}
\affiliation{
  \institution{Shandong University}
  \city{Qingdao}
  \country{China}
}
\email{majun@sdu.edu.cn}

\author{Maarten	de Rijke}
\affiliation{%
  \institution{University of Amsterdam}
  \city{Amsterdam}
  \country{The Netherlands}
}
\email{M.deRijke@uva.nl}

\author{Zhaochun Ren}
\authornote{Corresponding author.}
\affiliation{
  \institution{Leiden University}
  \city{Leiden}
  \country{The Netherlands}
}
\email{z.ren@liacs.leidenuniv.nl}

\renewcommand{\shortauthors}{Wu et al.}

\copyrightyear{2023}
\acmYear{2023}
\setcopyright{rightsretained}
\acmJournal{TOIS}

\begin{document}

\title[Learning Robust Sequential Recommenders through Confident Soft Labels]{Learning Robust Sequential Recommenders through \\ Confident Soft Labels}

\begin{abstract}
Sequential recommenders that are trained on implicit feedback are usually learned as a multi-class classification task through softmax-based loss functions on one-hot class labels. 
However, one-hot training labels are sparse and may lead to biased training and sub-optimal performance. 
Dense, soft labels have been shown to help improve recommendation performance. 
But how to generate high-quality and confident soft labels from noisy sequential interactions between users and items is still an open question.

We propose a new learning framework for sequential recommenders, CSRec, which
introduces \underline{\textbf{c}}onfident \underline{\textbf{s}}oft labels to provide robust guidance when learning from user-item interactions. 
CSRec contains a teacher module that generates high-quality and confident soft labels and a student module that acts as the target recommender and is trained on the combination of dense, soft labels and sparse, one-hot labels. 

We propose and compare three approaches to constructing the teacher module: 
\begin{enumerate*}[label=(\roman*)] 
\item model-level, 
\item data-level, and 
\item training-level. 
\end{enumerate*}
To evaluate the effectiveness and generalization ability of CSRec, we conduct experiments using various state-of-the-art sequential recommendation models as the target student module on four benchmark datasets. 
Our experimental results demonstrate that CSRec is effective in training better performing sequential recommenders.
\end{abstract}



\maketitle

\acresetall

\section{Introduction} 
 
Generating next-item recommendations from sequential implicit user feedback is a widely adopted way of training recommender systems.
It is common in scenarios like e-commerce~\citep{intro:e-commerce}, video platforms~\citep{intro:video-platforms}, and streaming music services~\citep{intro:music-recommemdation}. 
The problem of learning sequential recommenders based on implicit feedback can be formulated as a multi-class classification task where each candidate item corresponds to a class. 
Then, a list of recommendations is generated by choosing items with the highest classification logits. 

Deep neural networks have been widely used to address such classification tasks through softmax-based loss functions over one-hot class labels. 
We can view the items that a user has interacted with as being labeled as 1s, while all other items from the item catalogue are labeled as 0s~\citep{gru4rec,nextitnet,sasrec,narm}. 
Thus, the items a user has interacted with are interpreted as the user's positive preferences and are pushed towards higher classification logits in the training process; all other candidate items are assumed to capture negative user preferences. 
However, one-hot training labels are sparse and can easily be corrupted~\citep{intro:corrupted-one-hot,intro:noisy-label-effect}. 
For example, interactions with an item may only be due to presentation bias~\citep{intro:implicited-bias,intro:bias-in-rec} and an absent interaction may be attributed to user unawareness since the item may not have been exposed to this user~\citep{vardasbi-2020-inverse}. 
Hence, simply promoting high values for items labeled with a 1 and demoting other candidate items can lead to a misunderstanding of user preferences. 

\header{Soft labels}
Recent work \citep{distill-use-uniform-data, soft-rec} has shown that compared with sparse one-hot training labels, \emph{soft labels} can help to improve the recommendation performance. 
\emph{Soft} labels are labels that use class probabilities produced by models instead of the hard one-zero representation, i.e., one-hot labels.
They can be seen as a dense distribution over candidate items given the current sequence of items.
\citet{distill-use-uniform-data} propose a debiasing recommendation framework based on soft labels that originate from knowledge distillation of uniform exposure data.  
\citet{soft-rec} use popularity-based or user-based soft labels to train recommenders.

Motivated by \citep{distill-use-uniform-data,soft-rec}, we hypothesize that \emph{high-quality and confident soft labels can help to train more robust sequential recommenders}. 
Here, a soft label is \emph{confident} if the supervision signals that it provides reflect the real user preferences and have low variance.
And a sequential recommender is \emph{robust} if it is able to generate recommendations that lead to a positive user experience even if the training data is corrupted or contains noise.
Unfortunately, the uniform exposure data used in \citep{distill-use-uniform-data} will always be limited and expensive to collect since doing so may negatively affect the user experience by exposing irrelevant items.
And the soft labels used in \cite{soft-rec} can easily get biased or corrupted by the training data and the teacher model itself.

\begin{figure}
    \centering
    \begin{minipage}{\linewidth}
    \centering
        \captionsetup{labelformat=empty}
        \begin{minipage}{\linewidth}
        \centering
            \captionsetup[subfigure]{skip=-4pt}
            \subcaptionbox{Yelp}{\includegraphics[width = 0.45\linewidth]{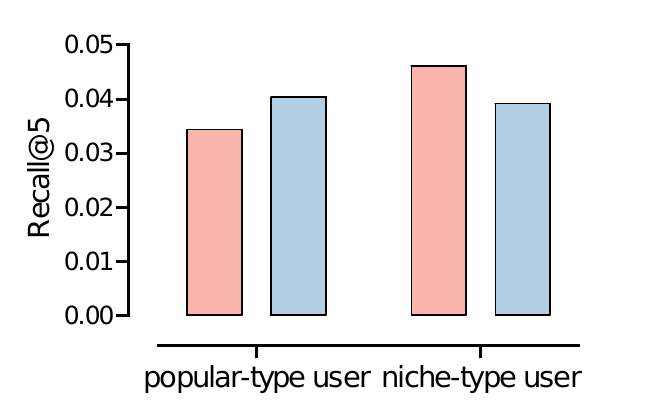}}
            \subcaptionbox{Electronics}{\includegraphics[width = 0.45\linewidth]{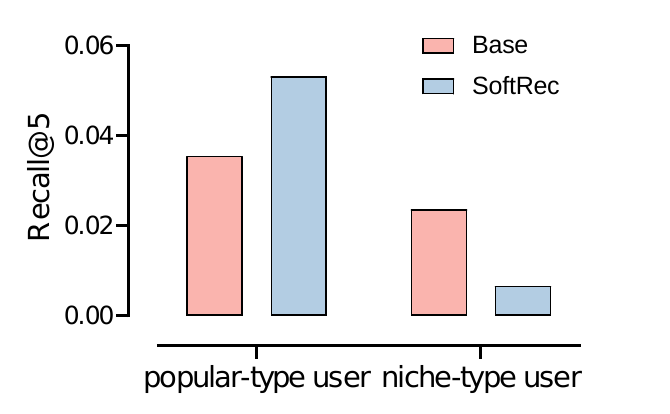}}
        \end{minipage}
    \end{minipage}
    \captionsetup{skip=4pt}
    \caption{Performance of Base recommendation model and SoftRec method \cite{soft-rec} on different groups of users.
The items are divided into two bins, 20\% and 80\%, based on the frequency.
We then split users based on which kind of items they choose as the next interactions, popular or niche items. The two user groups are referred to as \emph{popular-type}  and \emph{niche-type} users.
We evaluate the recommendation performance on each group separately.
We choose GRU4Rec~\citep{gru4rec} as the Base recommendation model and SoftRec~\citep{soft-rec} as the comparison method on the (a) Yelp and (b) Electronics dataset.
The niche-type users receive worse recommendation results after using the soft labels proposed in \cite{soft-rec}.}
    \label{fig:pop-bias-intro}
\end{figure}

As shown in Fig~\ref{fig:pop-bias-intro}, niche-type users, who choose niche items as their next selections, receive worse recommendation results after using the soft labels proposed in \cite{soft-rec}.
The bias of soft labels used in \cite{soft-rec} results in less exposure of niche items, which is not beneficial for both users and merchants in the long run.
Hence, how to create confident, soft labels from noisy sequential interactions between users and items remains an open question.

\header{Recommendations with confident soft labels}
We propose CSRec, a learning framework to train robust sequential recommenders from implicit user feedback. 
The core idea is to introduce confident, soft labels that complement the one-hot labels during the learning process. 
CSRec comprises a teacher module that generates confident, soft labels from noisy sequential interactions between users and items, and a student module that can be seen as the target recommender. 
We propose three methods for constructing the teacher module: 
\begin{enumerate*}[label=(\roman*)]
\item model-level, 
\item data-level, and 
\item training-level. 
\end{enumerate*}
These alternatives are motivated as follows:

\begin{itemize}%
    \item Recent research~\citep{Dodge2020FineTuningPL,method:quantized-analysis-of-different-init} has shown that different models, or even a set of instances of the same model initialized with different random seeds, introduce different kinds of bias into model outputs. Our \textbf{model-level} method constructs confident, soft labels from a multi-model ensemble to reduce the bias and variance from the model itself.
    \item Our \textbf{data-level} method reduces the bias or noisy signals coming from the data by exploiting sub-sampling procedures to feed the teacher models with different  subsets of the data. The confident, soft labels are, again, generated from a multi-data ensemble.
    \item The model-level and data-level methods use ensemble methods to obtain confident soft labels. Instead, the \textbf{training-level} method focuses on directly training a  teacher module in an end-to-end fashion. The key insight is to minimize the Kullback-Leibler (KL) divergence between predictions of two teacher models since confident, soft labels from different models should be consistent to provide more robust guidance.
\end{itemize}
\noindent%
Given confident, soft labels from the teacher module, the target student recommender is trained using a combination of dense, soft labels and sparse one-hot labels. 
Although knowledge distillation also utilizes soft labels, the motivation is quite different.
Knowledge distillation aims to compress a large neural network model into a relatively smaller model for more efficiency inference, while we aim at utilizing robust soft labels to generate robust recommendations from biased data, instead of speeding up the inference.
Besides, existing ensemble methods perform ensembles in both training and inference, while our method involves multiple models to generate confident soft labels in the training stage and only a single robust target recommender is used during inference, thus achieving higher recommendation efficiency.

\header{Experimental comparison}
To assess the effectiveness and generalization capability of recommender systems trained using the labels produced by CSRec, we conduct experiments on four benchmark datasets using different kinds of state-of-the-art sequential recommendation models as the target recommendation module: 
\begin{enumerate*}[label=(\roman*)] 
\item the recurrent neural network (RNN)-based GRU4Rec \citep{gru4rec}, 
\item the convolutional neural network (CNN)-based Nextitnet \citep{nextitnet}, 
\item the attention-based NARM \citep{narm}, and 
\item the self-attentive SASRec \citep{sasrec}. 
\end{enumerate*}
Experimental results demonstrate the effectiveness and generalization capability of the CSRec learning framework.

\header{Contributions}
Summarizing, the contributions of our work are:
\begin{itemize}%
    \item We propose CSRec, a learning framework to enhance implicit feedback-based sequential recommenders through a teacher module that provides confident, soft labels and a target student recommender trained on sparse, one-hot labels and dense, soft labels.
    \item We propose three methods for constructing the teacher module: \begin{enumerate*}[label=(\roman*)]\item a model-level method, \item a data-level method, and \item a training-level method,\end{enumerate*} all aimed at reducing the effect of bias and noisy interaction signals.
    \item We evaluate CSRec with four kinds of state-of-the-art deep learning-based sequential recommendation models and conduct experiments on four benchmark datasets. The results demonstrate the effectiveness of the CSRec learning framework.
\end{itemize}

\section{Related work}
We review related work on sequential recommendation and learning from soft labels.

\subsection{Sequential recommendation}
Early work on sequential recommendation mainly depends on factorization methods \citep{related:factorization-in-seq-rec,related:factorization-machine-for-seq-rec} or Markov chains \citep{related:markov-chain-in-seq-rec,related:markov-chain-and-similarity-model-in-seq-rec}. 
More recently, deep learning-based sequential recommendations have attracted attention due to their learning capacity. 
Deep learning-based sequential recommendation models can be categorized into 
\begin{enumerate*}[label=(\roman*)]
\item RNN-based models \citep{related:RNN-based-seq-rec,gru4rec,related:RNN-based-session-rec}, 
\item CNN-based models \citep{related:CNN-based-seq-rec}, and 
\item attention-based models \citep{related:attention-based-seq-rec,sasrec}. 
\end{enumerate*}
GRU4Rec \citep{gru4rec} is a representative RNN-based model that uses gated recurrent units to learn sequential signals. 
Caser~\citep{related:CNN-based-seq-rec} and Nextitnet~\citep{nextitnet} are CNN-based models that can capture skip signals in the interaction sequence. 
NARM~\citep{narm} introduces an attention mechanism to assign different degrees of importance to items that a user interacts with in the item sequence. 
SASRec~\citep{sasrec} uses self-attention and a transformer decoder \citep{transformer} to perform sequential recommendation. 
Unlike the causal decoding of SASRec, BERT4Rec \citep{related:bert4rec} uses the objective of masked language models to train the recommender.

Conventional training of sequential recommenders with implicit feedback uses the point-wise binary cross-entropy (BCE) loss or the pair-wise ranking loss (e.g., BPR \citep{related:BPR}).
Both need a negative sampling strategy to sample negative instances from missing interactions. 
The sampling size and distribution are likely to affect the recommendation performance \citep{related:sampling-matters}. 
Although there is work that focuses on non-sampling-based training methods, they are limited because of their high computational costs~\citep{related:non-sampling-methods-are-computational} or their shallow linear  models~\citep{related:non-sampling-are-restricted-to-linear-shallow-models}. 

Deep learning-based approaches to sequential recommendation are often formulated as multi-class classifiers where each candidate item corresponds to a class. The deep models are then trained through a softmax classification loss over the sparse one-hot class labels, in which items that users interacted with are labeled as 1s, and all other candidate items are labeled as 0s~\citep{gru4rec,nextitnet,sasrec,narm,related:bert4rec,related:GNN-based-seq-rec}. 
Such a solution assumes that the items that users interacted with indicate positive user preferences while all other candidate items reflect negative user preferences.
This assumption seldom holds for real-world cases~\citep{vardasbi-2020-inverse}. Interactions may be attributed to various kinds of presentation bias~\citep{intro:implicited-bias}; and in some cases, a lack of interaction may be attributed to the user's unawareness. Thus, sparse one-hot training labels can easily be corrupted and cannot capture unobserved user-item interactions. Solving the sequential recommendation task through sparse one-hot labels therefore leads to biased training and sub-optimal performance~\citep{intro:implicited-bias,distill-use-uniform-data,intro:corrupted-one-hot}. 

In this paper, we aim to learn sequential recommenders from implicit feedback data by using confident, soft labels in addition to sparse one-hot labels.

\subsection{Learning from soft labels}
\begin{figure*}
    \centering
	\subcaptionbox{Model-level CSRec}{\includegraphics[width = 0.32\textwidth]{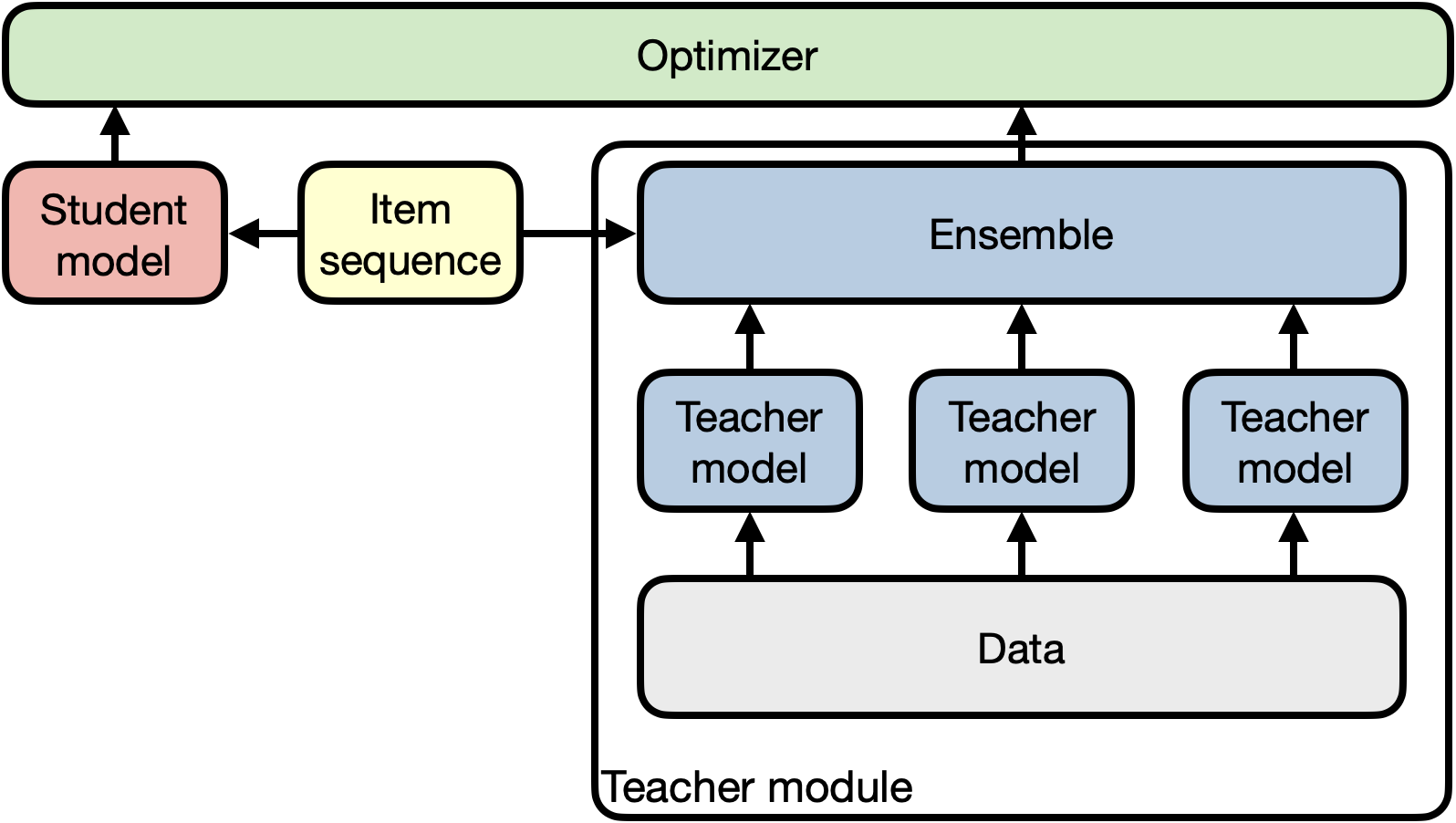}}
	\hfill
	\subcaptionbox{Data-level CSRec}{\includegraphics[width = 0.32\textwidth]{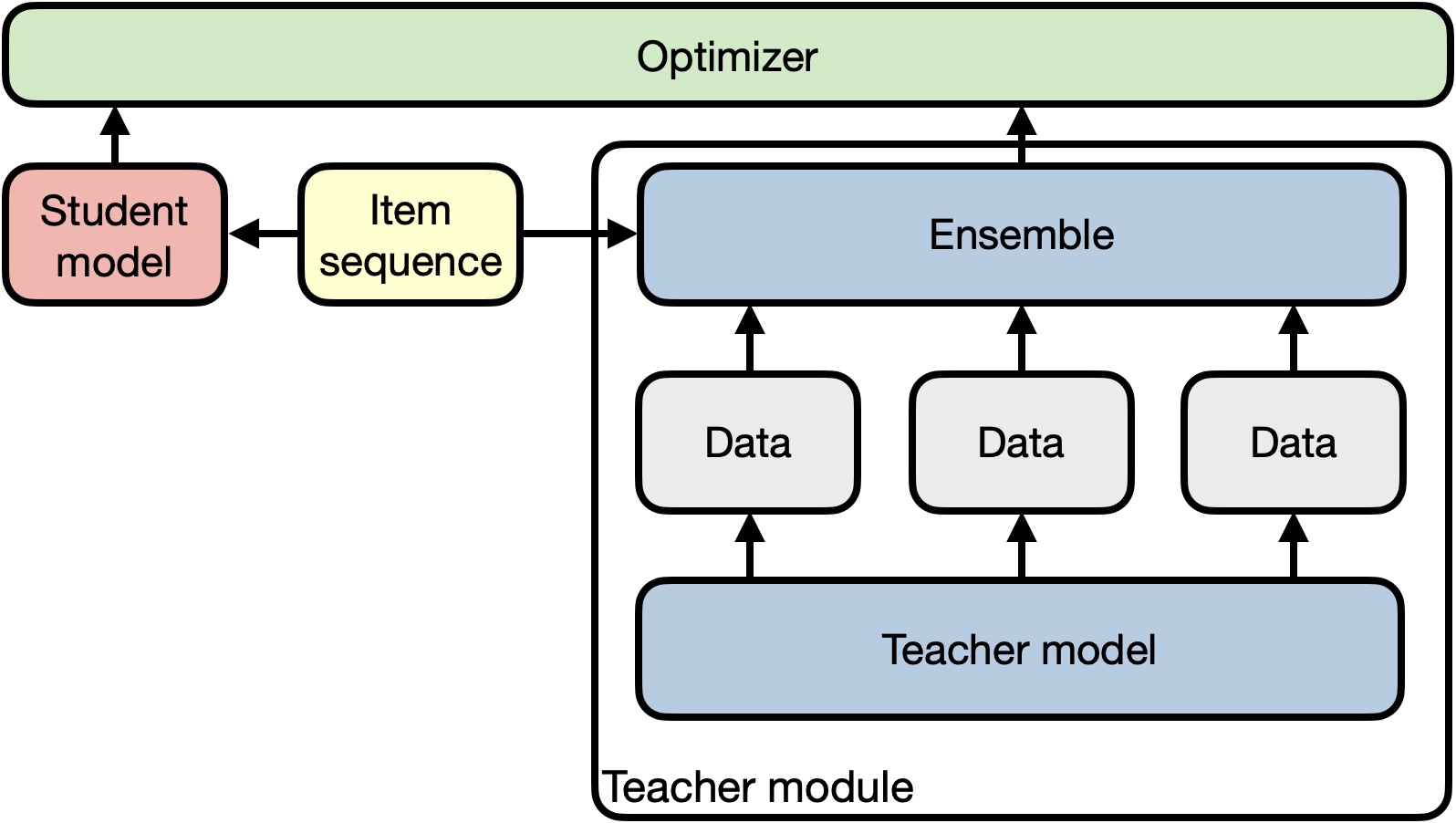}}
	\hfill
	\subcaptionbox{Training-level CSRec}{\includegraphics[width = 0.32\textwidth]{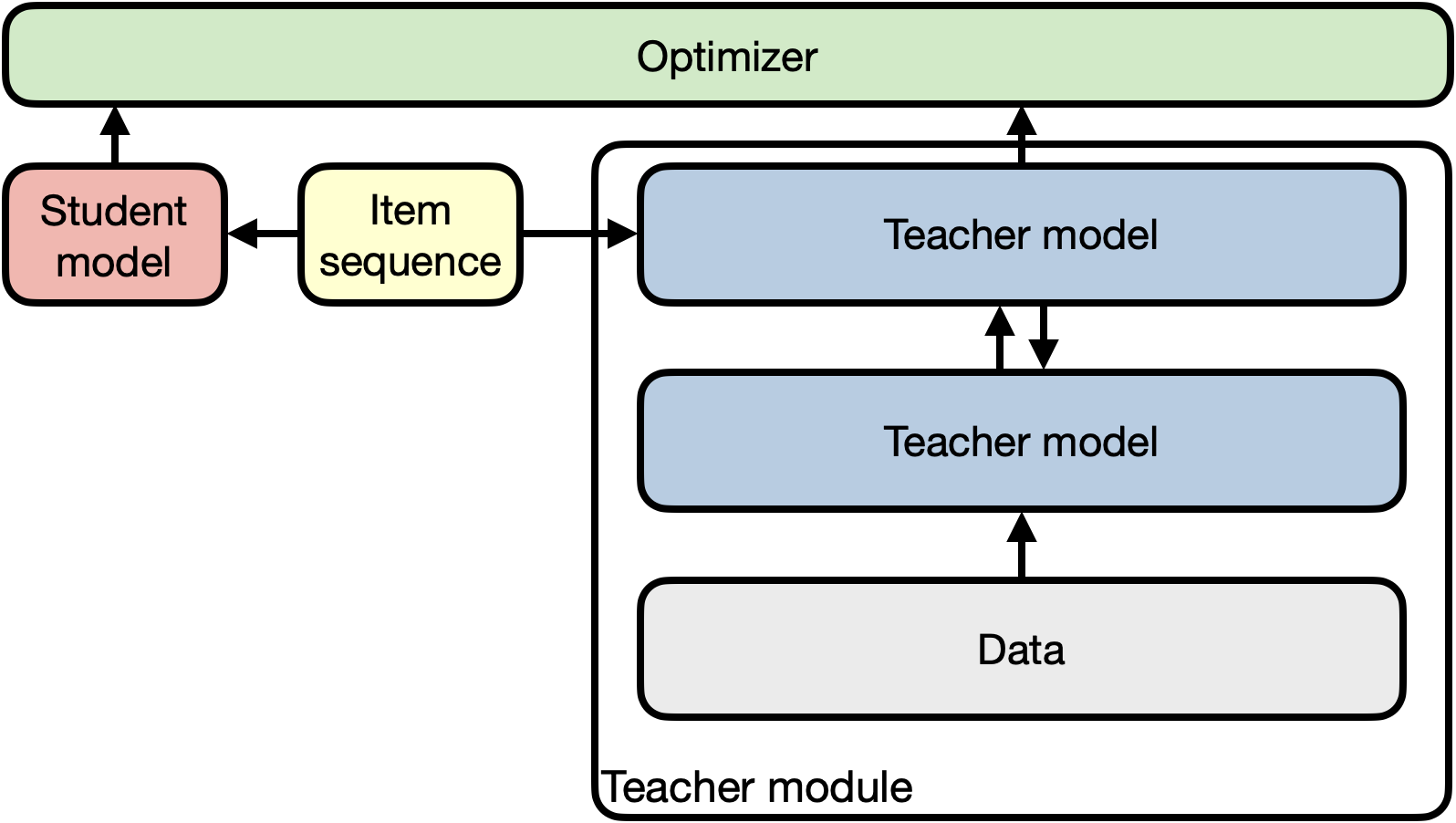}} 

    \caption{Framework of CSRec, with (a) model-level teachers, (b) data-level teachers, and (c) training-level teachers.}
    \label{fig:main-fig}
\end{figure*}

Soft labels have been successfully exploited in computer vision (CV) and natural language processing (NLP). 
Knowledge distillation \citep{intro:distill} is a widely adopted learning framework aimed at compressing large complex models into a simple model with a much smaller model size for faster model inference. Knowledge distillation extracts the hidden knowledge from large teacher networks in the form of soft labels to guide the training process of a much smaller student network.
Distillation using multi-models has also been investigated in CV and NLP. 
A common way of multi-model distillation is to use the average response from all teachers as part of the supervision signals for the student \citep{related:distill_survey,related:deep-mutual-learning,related:learing-from-multiple-teacher}. 
For example, \citet{related:learing-from-multiple-teacher} average the outputs of different teachers and reduce the dissimilarity between the student and teachers; \citet{related:adaptive_ensemble} use multi-objective optimization to determine the best direction that accommodates different teachers; and \citet{related:adaptive_multi_teacher} average the outputs of teachers with weights generated from themselves. 

Some recent work has exploited soft labels to improve recommendations. 
Even though the prediction from a well-trained teacher model contains beneficial  knowledge and may be helpful for building additional supervision for student models, naive distillation would bring unwanted bias and noisy signals. The root cause lies in the fact that, compared with the data in CV and NLP, the interaction data in recommendation contains more noise and biases.
\citet{Chen2022UnbiasedKD} find the biased student phenomenon during model compression using knowledge distillation.
They propose a grouping method to specifically tackle the popularity bias in the general recommendation system. 
Their method heavily relies on prior knowledge of the bias. 
Besides, their method focuses on improving the distillation process given biased soft labels, while our research is aimed at inferring more robust and confident soft labels, which works at different stages.
\citet{distill-use-uniform-data} propose a debiasing framework for recommendations based on soft labels learned from uniform interaction data. 
However, uniform interaction data is always limited and expensive to collect since it affects the user experience. 
\citet{soft-rec} use popularity-based and user-based soft labels to enhance sequential recommenders. Unfortunately, their generated soft labels can easily be corrupted by noisy implicit feedback.

To conclude, knowledge distillation in CV and NLP mainly focuses on using soft labels to perform model compression or transfer learning. 
In recommender systems, how to generate confident, soft labels from noisy implicit feedback from users for robust sequential recommendation is still an open research question.
We address this question head on and provide a new learning framework using soft labels to obtain robust sequential recommenders from implicit feedback.

\section{method}
\label{method}
In this section, we first share our notation and the task formulation. 
Then we describe three ways of constructing a teacher module that provides confident, dense, soft labels. 
Finally, we detail the training procedure of the target student recommender. 

\subsection{Notation and task formulation}
We focus on the task of learning sequential recommenders from implicit user feedback. 
We denote the user set and the item set as $\mathcal{U}$ and $\mathcal{I}$, respectively. 
Each user $u\in\mathcal{U}$ implicitly interacts with a sequence of items $\boldsymbol{s}_u=(i_1,i_2,\dots,i_{|\boldsymbol{s}_u|})$, sorted by time. 
Ideally, the next interacted item is the one with the highest probability in the real user preference distribution $P(i\mid \boldsymbol{s}_u)$.
However, the observed next item $i_{|\boldsymbol{s}_u|+1}$ can be corrupted by different kinds of noise (bias), which can be formulated as:
\begin{equation}
i_{|\boldsymbol{s}_u|+1}=\arg \max_{i\in\mathcal{I}} P(i\mid \boldsymbol{s}_u)+\epsilon,
\end{equation}
where $\epsilon$ denotes the noisy signal. 
The task is to train a target recommender $f$ that can approximate the real user preference as $P_f(i\mid\boldsymbol{s}_u)\approx P(i\mid\boldsymbol{s}_u)$. 
During model inference, recommendations can be generated as 
\begin{equation}
\tilde{i}=\arg \max_{i\in\mathcal{I}} f(\boldsymbol{s}_u),
\end{equation}
where $f(\boldsymbol{s}_u)$ denotes the output logits of recommender $f$. 
This task can be formulated as a multi-class classification problem with $\mathcal{I}$ being the candidate class set. 
Conventional methods that ignore the noisy signal $\epsilon$ use a softmax-based cross-entropy loss to train the recommender $f$ by minimizing the difference between $\tilde{i}$ and $i_{|\boldsymbol{s}_u|+1}$. 
However, due to the existence of $\epsilon$, $i_{|\boldsymbol{s}_u|+1}$ is not actually sampled from $P(i\mid\boldsymbol{s}_u)$, resulting in a discrepancy between the potentially misdirected learned distribution $P_f(i\mid\boldsymbol{s}_u)$ and the ideal unbiased distribution $P(i\mid\boldsymbol{s}_u)$.

To address the above issue, the proposed CSRec framework treats the target recommender $f$ as a student module and introduces a teacher module that consists of a set of models $\left\{g_1,g_2,\ldots,g_m\right\}$. The teacher module aims to generate confident, soft labels to 
\begin{enumerate*}[label=(\roman*)] 
\item provide dense supervision signals of large amounts of missing interactions; and
\item alleviate the effect of the noisy signal $\epsilon$. 
\end{enumerate*}
Then, the student model $f$ is trained on the combination of dense, soft labels and sparse data observation of $i_{|\boldsymbol{s}_u|+1}$. 
Fig~\ref{fig:main-fig} shows three alternative teacher modules for CSRec, which we detail next.

\subsection{Confident teacher module}
We describe the three alternative teacher modules for CSRec.

\subsubsection{Model-level teachers}
Recent research~\citep{Dodge2020FineTuningPL,method:quantized-analysis-of-different-init} shows that different models, or a set of the same models but with different random seeds, are able to introduce different types of bias into model outputs. 
This suggests that different models capture different aspects of $\epsilon$, as model-specific noisy signals $\epsilon_m$.
We propose to construct a confident teacher module from multiple models to benefit from this insight.
The key idea is to take the average outputs of multiple teacher models.
Thus, noisy signals $\epsilon_m$ along with their models are averaged, which gives us a more uniform error distribution and more robust predictions to generate soft labels.

We use a set of models $\{g_1,g_2,\ldots,g_m\}$ that are based on the same model architecture with the target student recommender $f$ but with different random seeds as the teacher module.
We train $\{g_1,\dots,g_m\}$ individually on the entire dataset using different training seeds. We use the softmax-based cross-entropy as the training loss function for the teacher models, denoted as $\ell_{ce}$. 
Then we take the average of the outputs to generate soft labels to guide the student.

The training process of model-level teachers is described in Algorithm~\ref{alg:post-model}. Model-level teachers are expected to alleviate bias or noise introduced by the model itself, i.e., $\epsilon_m$.

\begin{algorithm}[t]
\caption{Model-level teachers}
\label{alg:post-model}
\begin{algorithmic}[1]
\Require $\{g_1,g_2,\ldots,g_m\}$: teacher models;  $\eta$: learning rate; $\ell_{ce}(\cdot)$: loss for teacher models; $\theta(\cdot)$: parameters; $D$: training set
\Repeat
    \ForAll{$\boldsymbol{s}_u\in D$} 
        \State $\theta(g_k) \!\gets\! \theta(g_k) \!-\! \eta\cdot \nabla \ell_{ce}(\boldsymbol{s}_u,g_k,i_{|\boldsymbol{s}_u|+1})$\Comment{\mbox{}\!$k=1,2,\dots,m$}
    \EndFor
\Until{converged}

\ForAll{$\boldsymbol{s}_u\in D$}
    \State $\mathbf{e}_u \gets \sum g_k(\boldsymbol{s}_u)/ m$ \Comment{ensemble of $g_k$}
\EndFor
\State \textbf{return } $\mathbf{e}_u$
\end{algorithmic}
\end{algorithm}

\subsubsection{Data-level teachers}
Despite the noisy signal of $\epsilon_m$, which is reflected in the model-level perspective, there may also be data-level bias, $\epsilon_d$, as a part of the noisy signal $\epsilon$. We argue that $\epsilon_d$ is data-specific and has different distributions across different data subsets. 
As a result, a multi-data ensemble of models trained on different subsets of the data would help us alleviate the effect of $\epsilon_d$ and provide more robust guidance for the target student recommender. 

To this end, we propose to construct the confident teacher module from the data-level view. 
We perform subsampling with a uniform probability $p$ on the entire data and generate different subsets. 
Then, teacher models $\{g_1,g_2,\ldots,g_m\}$, which have the same model structure, are trained to fit different data subsets. Then we fuse the outputs of $\{g_1,g_2,\ldots,g_m\}$ to generate more robust teacher outputs. 

The data-level teacher’s algorithm is described in Algorithm~\ref{alg:post-data}.

\subsubsection{Training-level teachers}
Model- and data-level teachers use ensemble-based methods to fuse multi-source outputs, which can be referred to as \textbf{post-training} strategies. Training-level methods aim to directly learn a robust teacher module without a post-training ensemble.
The key idea is based on two assumptions:
\begin{enumerate}[leftmargin=*]
    \item confident, soft labels should be consistent with the latent real user preference; and
    \item confident, soft labels from two models should be consistent.
\end{enumerate}
Here, we introduce the main teacher model $g_1$ and a side teacher model $g_2$. We then describe how to directly train a robust $g_1$ with the help of $g_2$.

As discussed before, the output of $g_1$ and $g_2$ should be consistent with each other if they are expected to generate confident, soft labels. Therefore, given an item sequence $\mathbf{s}_u$, we aim to minimize their KL divergence as
\begin{equation}
\begin{split}
    \textrm{KL}(P_{g_2}(\cdot\mid\boldsymbol{s}_u)\|P_{g_1}(\cdot\mid\boldsymbol{s}_u))
    =\mathbb{E}_{j\sim P_{g_2}}[\log(P_{g_2}(j\mid\boldsymbol{s}_u))-\log(P_{g_1}(j\mid\boldsymbol{s}_u))],
\end{split}    
    \label{eq:pure_ml}
\end{equation}
where $j$ is a random variable representing the ideal user-preferred item that cannot be directly observed. What we can observe from the data is $i_{|\boldsymbol{s}_u|+1}$, which contains noise and bias. 
For simplicity, in the following description we will use $i$ to denote $i_{|\boldsymbol{s}_u|+1}$. Moreover, $\boldsymbol{s}_u$ will also be omitted occasionally and $P(\cdot\mid\boldsymbol{s}_u)$ is abbreviated as $P$ if necessary.

\begin{algorithm}[b]
\caption{Data-level teachers}
\label{alg:post-data}
\begin{algorithmic}[1]
\Require $\{g_1,g_2,\ldots,g_m\}$: teacher models; $\eta$: learning rate; $\ell_{ce}(\cdot)$: loss for teacher models; $\theta(\cdot)$: parameters; $D$: training dataset
\State randomly sample $p$ percent of $D$ for $m$ times as $D_1,\ldots,D_m$
\Repeat
    \ForAll{$\boldsymbol{s}_u\in D_k$} \Comment{$k=1,2,\ldots,m$}
        \State $\theta(g_k) \gets \theta(g_k) - \eta\cdot \nabla \ell_{ce}(\boldsymbol{s}_u,g_k,i_{|\boldsymbol{s}_u|+1})$
    \EndFor
\Until{converged}

    \ForAll{$\boldsymbol{s}_u\in D$}
        \State $\mathbf{e}_u \gets \sum{g_k}(\mathbf{s}_u) / m$ \Comment{ensemble of $g_k$}
    \EndFor

\State \textbf{return } $\mathbf{e}_u$
\end{algorithmic}
\end{algorithm}

According to Bayes' Theorem and the discussed assumption, we have
\begin{equation}
    P_{g_1}(j\mid\boldsymbol{s}_u)\sim P(j\mid\boldsymbol{s}_u) = \frac{P(i\mid\boldsymbol{s}_u)P(j\mid i, \boldsymbol{s}_u)}{P(i\mid j,\boldsymbol{s}_u)}.
    \label{eq:bayes}
\end{equation}
Substituting Eq.~\ref{eq:bayes} into Eq.~\ref{eq:pure_ml}, we have
\begin{align}
    \textrm{KL}(P_{g_2}\|P_{g_1})
    &{}\approx \mathbb{E}_{j\sim P_{g_2}}\left[\log(P_{g_2}(j\mid \boldsymbol{s}_u))-\log\frac{P(i\mid \boldsymbol{s}_u)P(j\mid i, \boldsymbol{s}_u)}{P(i\mid j,\boldsymbol{s}_u)}\right]\nonumber\\
    &{}=\textrm{KL}(P_{g_2}\| P)-\log(P(i\mid \boldsymbol{s}_u))+\mathbb{E}_{j\sim  P_{g_2}}[\log P(i\mid j,\boldsymbol{s}_u)].
    \label{eq:derivation}
\end{align}
If we rearrange Eq.~\ref{eq:derivation}, we have
\begin{equation}
\begin{split}    
    \textrm{KL}(P_{g_2}\| P)-\log(P(i\mid \boldsymbol{s}_u))
    \approx\textrm{KL}(P_{g_2}\| P_{g_1})-\mathbb{E}_{j\sim  P_{g_2}}[\log P(i\mid j,\boldsymbol{s}_u)].
\end{split}    
    \label{eq:rearranged}
\end{equation}
Due to the fact that $\textrm{KL}(P_{g_2}\| P)\ge 0$, the right-hand side of Eq.~\ref{eq:rearranged} is an approximate upper bound of the negative logarithm likelihood. We can regard $\mathbb{E}_{j\sim  P_{g_2}}[\log P(i\mid j,\boldsymbol{s}_u)]$ as a regularization term to adjust the agreement across $ P_{g_1}$, $ P_{g_2}$ and $P$.

Through a similar derivation, we also have
\begin{equation}
    \begin{split}
    \textrm{KL}(P_{g_1}\| P)-\log(P(i\mid \boldsymbol{s}_u))
    \approx\textrm{KL}(P_{g_1}\| P_{g_2})-\mathbb{E}_{j\sim  P_{g_1}}[\log P(i\mid j,\boldsymbol{s}_u)].
    \end{split}
    \label{eq:rearranged2}
\end{equation}
Combining the right-hand sides of both Eq.~\ref{eq:rearranged} and Eq.~\ref{eq:rearranged2}, we obtain the following regularization term:
\begin{equation}
\begin{split}
    \ell_{reg}=\;&\alpha \left\{\textrm{KL}(P_{g_2}\| P_{g_1})-\mathbb{E}_{j\sim P_{g_2}}[\log P(i\mid j,\boldsymbol{s}_u)]\right\} \\
    &+ (1-\alpha)\left\{\textrm{KL}(P_{g_1}\| P_{g_2})-\mathbb{E}_{j\sim P_{g_1}}[\log P(i\mid j,\boldsymbol{s}_u)]\right\}.
\end{split}
\end{equation}
Since our goal is to train the main teacher $g_1$ with the help of side teacher $g_2$, the term $\mathbb{E}_{j\sim P_{g_2}}[P(i\mid j,\boldsymbol{s}_u)]$ has no contribution to the training of $g_1$, so we omit it and change our regularization function to
\begin{equation}
\begin{split}
    \ell_1=\alpha \textrm{KL}(P_{g_2}\| P_{g_1}) + (1-\alpha)\textrm{KL}(P_{g_1}\| P_{g_2})
    -\mathbb{E}_{j\sim P_{g_1}}[\log P(i\mid j,\boldsymbol{s}_u)].
\end{split}
\end{equation}
In practice, for simplicity, we assume that the observation of $i$ given the real user-preferred item $j$ is conditionally independent with the item sequence $\mathbf{s}_u$. 
This is a reasonable assumption since the correlation between the observed item $i$ and the ideal item $j$ mainly depends on the noisy signals. 
As a result, $P(i\mid j,\boldsymbol{s}_u)$ can be approximated through an auxiliary model $h$ as 
\begin{equation}
h(i,j)\approx P(i|j) \approx P(i\mid j,\boldsymbol{s}_u).
\end{equation}
A concise solution to learn $h$ is to use matrix factorization and factorize $h(i\mid j)$ as
\begin{equation}
[h]_{|\mathcal{I}|\cdot|\mathcal{I}|}= \mathcal{M}_{|\mathcal{I}|\cdot d}\cdot \mathcal{N}_{d\cdot|\mathcal{I}|},
\end{equation}
where $d\ll\mathcal{I}$; $[h]$ can be viewed as a global noise matrix. $\mathcal{M}$ and $\mathcal{N}$ are its low-rank factorization matrices.

To sum up, given an item sequence $\mathbf{s}_u$ and the observed next item $i$ as the label, we have the training-level robust regularization
\begin{equation}
\begin{split}
\ell_r=\alpha \textrm{KL}(P_{g_2}\| P_{g_1})+(1-\alpha) \textrm{KL}( P_{g_1}\| P_{g_2})
- \sum_{j\in \mathcal{I}}\log(h_{i,j}) P_{g_1}(j\mid \mathbf{s}_u).
\end{split}
\label{eq:dvae_loss}
\end{equation}
Finally, we combine the robust loss $\ell_r$ with the softmax cross-entropy loss (i.e., $\ell_{ce}$) as the final loss function to construct the training-level confident teacher module:
\begin{equation}
\ell_t=\ell_r+\ell_{ce}.
\end{equation}
We pretrain the side teacher model $g_2$ on a sub-sampled data split to enhance the learning stability and further introduce data-level robustness. The algorithmic process of training-level teachers is described in Algorithm~\ref{alg:in-training}.

\begin{algorithm}[t]
\caption{Training-level teachers}
\label{alg:in-training}
\begin{algorithmic}[1]
\Require $g_1$, $g_2$: two teacher models; $\eta$: learning rate; $\ell_{ce}(\cdot)$: loss function for the side teacher $g_2$;
$\ell_{t}(\cdot)$: loss function for the main teacher $g_1$; $\theta(\cdot)$: parameters; $D$: training dataset

\State randomly sample $p$ percent data $D^\prime$ from $D$.

\Repeat
    \ForAll{$\boldsymbol{s}_u\in D^\prime$}
        \State $\theta(g_2)\gets\theta(g_2)-\eta\cdot\nabla \ell_{ce}(\boldsymbol{s}_u,g_2,i_{|\boldsymbol{s}_u|+1})$ \Comment{pretrain $g_2$}
    \EndFor
\Until{converged}

\Repeat
    \ForAll{$\boldsymbol{s}_u\in D$}
        \State $\theta(g_1)\gets\theta(g_1)-\eta\cdot\nabla \ell_t(\boldsymbol{s}_u,g_1,i_{|\boldsymbol{s}_u|+1})$ \Comment{train $g_1$}
    \EndFor
\Until{converged}

\ForAll{$\boldsymbol{s}_u\in D$}
        \State $\mathbf{e}_u \gets {g_1}(\mathbf{s}_u) $
\EndFor

\State \textbf{return } $\mathbf{e}_u$
\end{algorithmic}
\end{algorithm}

\subsection{Learning of student recommenders}
\label{sec:distill}
Given the input sequence $\mathbf{s}_u$ and dense logits $\mathbf{e}_u$ (see the output of Algorithm 1, 2, and 3) generated by the confident teacher module, we define the soft labels for the student recommender $f$ as
\begin{equation}
\mathbf{r}_{u}=\frac{1}{2}(\textrm{softmax}(\mathbf{e}_u/\mathcal{T})+\textrm{onehot}(i_{|\boldsymbol{s}_u|+1})),
\end{equation}
where $\mathcal{T}$ is a temperature smoothening the soft label distribution. A larger $\mathcal{T}$ indicates smoother soft labels. When $\mathcal{T}\rightarrow \infty$, the soft labels act as naive label smoothing.

Finally, the training loss for the target student recommender $f$ is defined as 
\begin{equation}
\ell_s=(1-\beta)\ell_{\textrm{ce}}(\boldsymbol{s}_u,f,i_{|\boldsymbol{s}_u|+1})+\beta \textrm{KL}(P_f(\cdot\mid \boldsymbol{s}_u)\|\mathbf{r}_u).
\label{eq:soft_loss}
\end{equation}
During inference, we use the student model $f$ to generate the list of recommended items by selecting the top-$n$ items with the highest classification logits.

\subsection{Summary and remarks}
We have proposed three strategies to learn a confident teacher module to generate soft labels. All three methods leverage the collaboration of multiple teachers but from different perspectives and in different stages of training. 
The model-level and data-level strategies use collaboration after the training of teachers, and the training of multiple teacher models is processed individually. Averaging the outputs is an effective way of reducing errors and making more robust soft labels. In contrast, the training-level strategy uses the collaboration between the main teacher model and a side teacher model during the training process. The main teacher model is directly trained to be robust without post-training infusion.
In the training-level strategy, we use a pre-trained $g_2$ instead of training both models together because a pre-trained model can dramatically accelerate the model collaboration.
Note that we use the sub-sampled data split during the pre-training of the side teacher model $g_2$. This design can help the main teacher model $g_1$ to benefit from the data-level randomness to generate more robust outputs.

\section{Experiments}
In this section, we describe our experimental setup for assessing the effectiveness of our proposed training method. We focus on the following research questions:
 \begin{enumerate}[label=(RQ\arabic*)]
    \item What is the overall recommendation performance of the proposed CSRec learning framework?
    \item Does CSRec help to generate more robust sequential recommendations from implicit user feedback? 
    \item How does the design of the teacher module affect the student performance?
\end{enumerate}

\subsection{Experimental settings}
\subsubsection{Datasets} We conduct experiments on four datasets: \begin{enumerate*}[label=(\roman*)]
\item Last.\-FM,\footnote{\url{https://grouplens.org/datasets/hetrec-2011/}} 
\item Yelp,\footnote{\url{https://www.yelp.com/dataset}} 
\item Amazon Electronics, and 
\item Amazon Movies and TV.\footnote{\url{http://jmcauley.ucsd.edu/data/amazon/}} 
\end{enumerate*}
Table \ref{tab:dataset} shows the dataset statistics. 
We remove users and items with less than five interactions for all datasets. Since we consider sequential recommendation tasks, we split each user's interactions into sequences with fixed lengths. If the session is too short, we add padding tokens. If the session is longer, we cut it into several sub-sessions.

\subsubsection{Evaluation protocols}
To evaluate the recommendation performance, we adopt the \textit{leave-one-out} evaluation procedure. The last item in a sequence is left as the test sample, while the one but last item is used for validation. The remaining interactions are used as the training set. We use the full item set as the candidate set when performing the ranking.
For evaluation, we use two ranking-based metrics: 
\begin{enumerate*}[label=(\roman*)]
\item \textit{Recall} and 
\item \textit{Normalized Discounted Cumulative Gain} (NDCG).
\end{enumerate*}
Recall@$n$ measures whether the ground-truth item occurs in the top-$n$ positions of the list of recommendations. NDCG is a weighted version that attaches higher importance to top positions. 

To evaluate the robustness of the recommen\-ders (i.e., whether the recommendation can lead to higher user ratings given the noisy implicit feedback), we also introduce filtered versions of these metrics on the two Amazon datasets (since those contain rating information). 
\label{exp:filtered-metrics}
The original Recall is defined as $\textrm{Recall}@n=\frac{1}{|\mathcal{U}|}\sum_{u\in \mathcal{U}}\mathbbm{1}(rank_u\leq n)$, where $\textrm{rank}_u$ denotes the rank of the ground-truth item and $\mathbbm{1}(\cdot)$ is the indicator function. 
For the \emph{filtered version}, which we denote as $\textrm{Recall}^+$, we have
\begin{equation}
    \textrm{Recall}^+@n=\frac{1}{\sum_{u\in\mathcal{U}} \mathbbm{1}(r_u \geq \delta)}\sum_{u\in \mathcal{U}}\mathbbm{1}(r_u\geq \delta \wedge rank_u\leq n),
\end{equation}
where $r_u$ is the rating of the ground-truth item. $\delta$ is a threshold value set to 4 (ratings range from 0 to 5). Such a filtered metric measures whether the recommended items can lead to real positive user preferences.
Similarly, we have a filtered version NDCG$^+$ of the NDCG metric. Note that ratings are only used in the evaluation stage to verify the robustness. For training, we only use binary implicit user feedback (i.e., items that a user interacted with are labeled as positive, and the others are missing interactions). 

\begin{table}
\caption{Statistics of datasets.}
\begin{tabular}{l r r r c c} 
 \toprule
 Dataset & \#users & \#items & \#interactions & length \\
 \midrule
 Last.FM  & 1,090 & 3,646 & 52,551 & 20 \\ %
 Yelp  & 30,431 & 20,033 & 316,354 & 10 \\ %
 Electronics  &  83,427 & 29,351 & 684,449 & 10\\ %
 Movies \& TV  & 47,966 & 21,035 & 625,813 & 10\\
 \bottomrule
\end{tabular}
\label{tab:dataset}
\end{table}

\subsubsection{Baselines}
To assess the effectiveness of training recommen\-ders with CSRec, we use the following models as the student model described in Section~\ref{method}:
\begin{itemize}%
    \item GRU4Rec~\citep{gru4rec} uses a GRU to model sequential user behavior, and we use an improved version by \citet{gru4rec-full}.
    \item NextItNet~\citep{nextitnet} is a simple yet effective CNN-based model that is capable of learning both short- and long-range dependencies.
    \item SASRec~\citep{sasrec} is based on the transformer decoder~\citep{transformer}, which uses multi-head self-attention to capture user preference.
    \item NARM~\citep{narm} uses an attention mechanism to capture the intention of users; then, it computes the recommendation score with a bi-linear matching based on the latent representation.
\end{itemize}
Each student recommendation model is trained with the following methods:
\begin{itemize}%
    \item \textbf{Base} denotes the standard training framework by optimizing the cross-entropy between the output and the sparse one-hot label;
    \item \textbf{SoftRec} denotes the item-based method proposed by \citet{soft-rec}, which uses a popularity-based teacher model to generate soft labels;
    \item \textbf{CSRec-M} is our proposed framework that uses the model-level teacher module to produce the confident, soft labels;
    \item \textbf{CSRec-D} is our proposed framework that uses the data-level teachers to provide confident, soft labels; and
    \item \textbf{CSRec-T} is our proposed framework that uses the training-level teachers to generate confident, soft labels.
\end{itemize}

\subsubsection{Implementation details}
We use the implementations from Recbole~\citep{exp:recbole},\footnote{\url{https://github.com/RUCAIBox/RecBole}} for the four baseline models listed above. 
The embedding size of all models, including $d$ in the training-level method, is fixed to $64$ for a fair comparison.
Both $\alpha$ and $\beta$ in Eq.~\ref{eq:dvae_loss} and Eq.~\ref{eq:soft_loss} are chosen in $\{0.25, 0.5, 0.75\}$, and the temperature $\mathcal{T}$ in Eq.~\ref{eq:soft_loss} is chosen in $\{1, 3, 6, 9\}$. 
For CSRec-M and CSRec-D, the number of teachers (i.e., $m$) is set to 2. For CSRec-D, the sub-sampling ratio $p$ is set to 0.8.
For GRU4Rec, the number of layers is set to $2$ on all datasets. The dropout rate is $0.5$.
For NextItNet, the kernel size is set to 3, the block number is $5$ on four datasets, and the dilations are set to $1$ after tuning.
For SASRec, the number of layers is set to $2$, the head number is $2$, and the dropout rate is $0.3$.
For NARM, the hidden size is $128$, and the number of layers is 1. %
We use Adam~\citep{Adam} as the training optimizer, setting the learning rate to $0.001$.
For  CSRec, we use the same model architecture for the teacher models $g$ as for the student models $f$ for a fair comparison.\footnote{The code and data used are available at \url{https://github.com/Furyton/CSRec/}.}

\begin{table*}
        \caption{Comparison of the top-$n$ recommendation performance of different models ($n=10$) on four datasets. RC is short for Recall@10. NG is short for NDCG@10.  \textbf{Boldface} denotes the highest score. $**$ and $*$ indicate significant improvements over the corresponding SoftRec baseline ($p < 0.05$ and $p<0.1$, respectively).}
        \centering
        \small
    
    \begin{tabular}{p{1.2cm}p{1.2cm} cc cc cc cc}
    
        \toprule
        & & \multicolumn{2}{c}{GRU4Rec}&\multicolumn{2}{c}{NextItNet}&\multicolumn{2}{c}{NARM}&\multicolumn{2}{c}{SASRec}
        \\
        \cmidrule(lr){3-4}\cmidrule(lr){5-6}\cmidrule(lr){7-8}\cmidrule(lr){9-10}
         \bf Dataset & \bf Method & RC(\%) & NG(\%) &RC(\%) & NG(\%)&RC(\%) & NG(\%)&RC(\%) & NG(\%)\\
        \hline
        \multirow{5}*{Last.FM} & 
        Base & 16.683 & 10.081 & 15.456 & 11.564 & 20.260 & \textbf{13.204} & 20.350 & 11.432 \\
        & SoftRec & 17.427 & 10.624 & 15.473 & 11.826 & 20.486 & 12.826 & 20.746 & 11.951 \\
        &CSRec-M & 18.165\rlap{*} & 10.678 & \textbf{16.466}\rlap{*} & \textbf{12.472}\rlap{**} & \textbf{20.819}\rlap{*} & 13.050 & 20.874 & 11.976 \\
        &CSRec-D & \textbf{18.607}\rlap{*} & \textbf{10.992}\rlap{*} & 16.201\rlap{*} & 12.055 & 20.739\rlap{*} & 12.960 & \textbf{21.131}\rlap{*} & \textbf{12.256}\rlap{*} \\
        &CSRec-T & 18.180\rlap{**} & 10.764\rlap{**} & 15.681 & 12.007 & 20.697 & 12.875 & 21.047 & 12.009 \\
        \midrule
        \multirow{5}*{Yelp} & 
        Base & 5.812 & 3.512 & 8.077 & 5.742 & 8.587 & 5.641 & 8.523 & 5.329\\
        & SoftRec & 6.108 & 3.584 & 8.303 & 5.923 & 8.108 & 5.024 & 9.459 & 5.882\\
        &CSRec-M & 6.405\rlap{**} & 3.822\rlap{**} & 8.587\rlap{*} & 6.175\rlap{*} & \textbf{8.900}\rlap{**} & \textbf{5.778}\rlap{**} & \textbf{9.485}\rlap{*} & \textbf{5.902}\rlap{*} \\
        &CSRec-D & 6.237 & 3.646 & 8.501 & 6.101 & 8.723\rlap{**} & 5.489\rlap{**} & 9.484 & 5.898\rlap{**} \\
        &CSRec-T & \textbf{6.526}\rlap{**} & \textbf{3.894}\rlap{**} & \textbf{8.589}\rlap{*} & \textbf{6.177}\rlap{*} & 8.700\rlap{**} & 5.530\rlap{**} & 9.421 & 5.888\\
        \midrule
        \multirow{5}*{Electronics} & 
        Base & 5.178 & 2.855 & 4.900 & 3.217 & 5.211 & 3.051 & 6.932 & 4.000 \\
        & SoftRec & 5.431 & 3.046 & 5.204 & 3.429 & 5.490 & 3.197 & 6.963 & 4.081 \\
        &CSRec-M & 5.545\rlap{**} & 3.132\rlap{**} & \textbf{5.666}\rlap{**} & \textbf{3.789}\rlap{**} & 5.787\rlap{**} & 3.432\rlap{**} & 6.737 & 3.944 \\
        &CSRec-D & 5.515\rlap{**} & 3.104\rlap{**} & 5.308\rlap{*} & 3.507\rlap{**} & 5.846\rlap{**} & 3.486\rlap{**} & 6.729\rlap{*} & 3.923 \\
        &CSRec-T & \textbf{5.595}\rlap{*} & \textbf{3.168}\rlap{**} & 5.356\rlap{**} & 3.583\rlap{**} & \textbf{5.905}\rlap{**} & \textbf{3.523}\rlap{**} & \textbf{7.001}\rlap{*} & \textbf{4.107}\rlap{*}\\
        \midrule
        \multirow{5}{*}{\begin{tabular}{@{}l@{}}Movies \& \\TV\end{tabular}} & 
        Base & 8.413 & 4.772 & 7.377 & 4.448 & 9.383 & 5.514 & 11.191 & 6.716 \\
        & SoftRec & 9.541 & 5.365 & 8.228 & 4.976 & 9.402 & 5.451 & 10.908 & 6.547 \\
        &CSRec-M & 9.638 & 5.521\rlap{*} & \textbf{8.744}\rlap{**} & \textbf{5.302}\rlap{**} & \textbf{9.999}\rlap{**} & \textbf{5.880}\rlap{**} & \textbf{11.198}\rlap{**} & 6.701\rlap{**} \\
        &CSRec-D & 9.458 & 5.426\rlap{**} & 8.367 & 5.029 & 9.754\rlap{**} & 5.718\rlap{**} & 11.195\rlap{*} & \textbf{6.717}\rlap{*} \\
        &CSRec-T & \textbf{9.704}\rlap{**} & \textbf{5.522}\rlap{**} & 8.461\rlap{**} & 5.116\rlap{**} & 9.876\rlap{**} & 5.777\rlap{**} & 11.195\rlap{*} & 6.703\\
       \bottomrule
        \end{tabular}
        \label{tab:main_result}
    \end{table*}

\section{Results and analysis}
\subsection{Overall performance comparison (RQ1)}

Table~\ref{tab:main_result} compares the recommendation performance of all models under different training regimes. 
Training with the three proposed CSRec methods (i.e., CSRec-M, CSRec-D, and CSRec-T) consistently leads to improved recommendation performance on four datasets and four student models, especially the model-level and the training-level methods, which achieve significant improvements over the SoftRec method in most cases.

If we break down the results by dataset, we see that training with CSRec-D sometimes leads to higher recommendation performance than training with the other two proposed methods on the Last.FM dataset. 
On the other three datasets, training with either CSRec-M or CSRec-T results in the highest scoring recommendation performance in most cases.
The reason could be that the interactions in Last.FM are much denser than in the other three datasets.
As the data-level teachers drop parts of the samples, more variations and randomness are produced so that CSRec-D can capture a more robust signal from the data.
Training with CSRec-T consistently leads to good recommendation performance on the Yelp and Electronics datasets, which are both relatively sparse.
We thus infer that the CSRec-D is suitable for training on dense datasets while CSRec-T fits training on sparse datasets. In addition, CSRec-M can be used in more general cases.

If we break down the results by student model, we see that RNN-based and CNN-based models, i.e., GRU4Rec and NextItNet, enjoy the most notable improvements when being trained with CSRec, while the SASRec model equipped with a self-attention mechanism has a relatively small improvement. Recent research \citep{exp:transformer-is-already-robust,exp:transformer-robust-ability} has shown that pre-trained transformers are far more effective at handling noisy examples than CNN or RNN-based models.
Similar conclusions have been found concerning computer vision~\citep{exp:transformer-robust-in-CV-literature}. 
However, in most cases, training with the proposed CSRec still leads to significant improvements in the performance of SASRec.

In summary, the proposed learning framework CSRec helps to effectively and significantly improve the performance of diverse recommenders compared with the normal training regime and SoftRec. We can choose different teacher modules based on different dataset characteristics to achieve the best performance.

\subsection{Robustness performance (RQ2)}
\label{sec:RQ2}
\subsubsection{Evaluation on the real positive user preference}
We use the filtered evaluation metrics on the two Amazon datasets to determine whether training with CSRec yields recommenders that lead to real positive user preferences (i.e., higher ratings). 
The  results are shown in Table~\ref{tab:robust_result}. 
The results demonstrate that our proposed training methods outperform the base training and SoftRec on the two Amazon datasets. 
Most CSRec results achieve significant improvements, indicating that our methods can effectively capture real user preferences even though the training dataset contains noisy training instances (i.e., items a user interacted with but with low ratings). Moreover, the high performance resulting from training with CSRec-M and CSRec-T also shows the generalization capability of our proposed methods.

\subsubsection{Effect on reducing the popularity bias}
\begin{figure}
    \centering
    \begin{minipage}{\linewidth}
    \centering
        \captionsetup{labelformat=empty}
        \begin{minipage}{\linewidth}
        \centering
            \captionsetup[subfigure]{skip=-4pt}
            \subcaptionbox{Yelp}{\includegraphics[width = 0.45\linewidth]{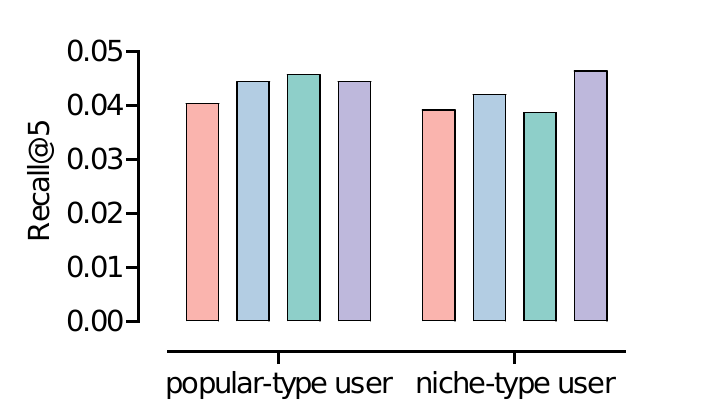}}
            \subcaptionbox{Electronics}{\includegraphics[width = 0.45\linewidth]{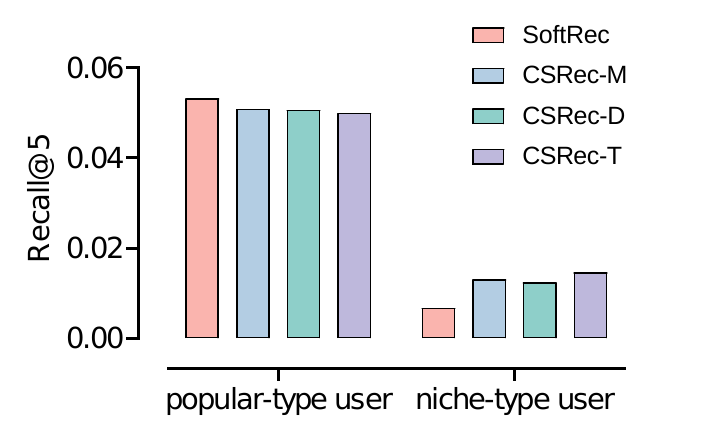}}
        \end{minipage}
    \end{minipage}
    \captionsetup{skip=4pt}
    \caption{Performance of SoftRec and the three proposed training methods, i.e., CSRec-M, CSRec-D, and CSRec-T, on different groups of users. Similar settings to Fig.~\ref{fig:pop-bias-intro} are adopted. We choose GRU4Rec~\citep{gru4rec} as the base model on the (a) Yelp and (b) Electronics dataset.}
    \label{fig:abla-pop-bias}
\end{figure}

 We examine the popularity bias of the student module in Fig~\ref{fig:abla-pop-bias}.
The recommendation performance is improved on the tail items with a small cost of decreasing on the most popular ones.
On the Yelp dataset, we even have large improvements for both popular-type and niche-type users.
This indicates that our training methods can alleviate popularity bias and focus more on niche items, which is beneficial for the long-term profits of service providers.

For a deeper analysis, popularity bias is not always harmful for the recommender according to~\citet{PopularityBiasIsNotEvil}, popular items sometimes do reflect the general user interests.
A single teacher model trained separately (e.g., SoftRec \cite{soft-rec}) can contain both harmful biases and beneficial general interests, and thus enhance the popularity bias.
However, the proposed CSRec can mitigate  harmful bias through the collaboration of multiple teacher modules, and it keeps the beneficial general interests. The reason is that the beneficial general user interests should be consistent among all the models and thus can be reserved during model collaboration. 

\subsubsection{Effect of sequence length and data sparsity}
\begin{figure}
    \centering
    \begin{minipage}{\linewidth}
    \centering
        \captionsetup{labelformat=empty}
        \begin{minipage}{\linewidth}
        \centering
            \captionsetup[subfigure]{skip=-4pt}
            \subcaptionbox{Effect of sequence length.\label{fig:abla-effect-of-length}}{\includegraphics[width = 0.49\linewidth]{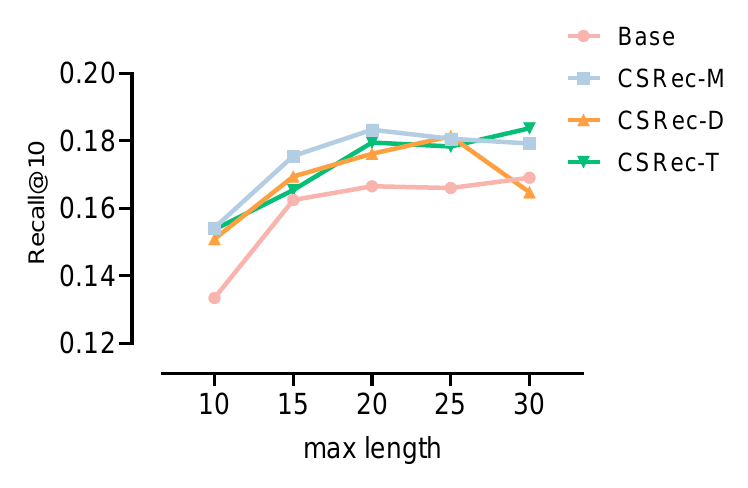}}
            \subcaptionbox{Effect of dataset sparsity.\label{fig:different-sparsity}}{\includegraphics[width = 0.49\linewidth]{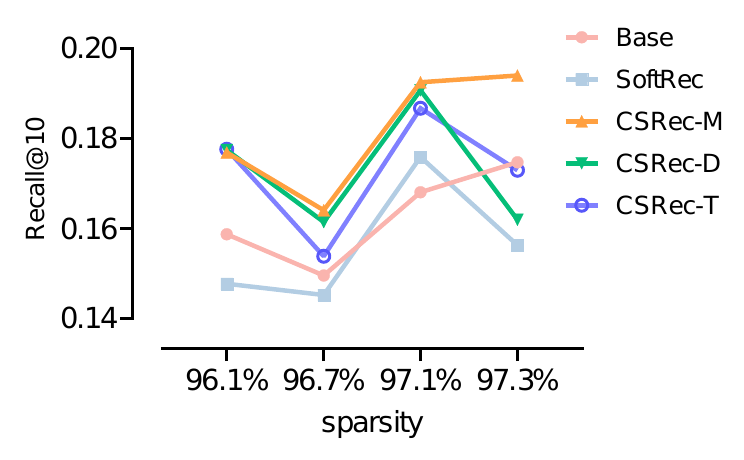}}
        \end{minipage}
    \end{minipage}
    \captionsetup{skip=4pt}
    \caption{Performance of Base model and the three proposed training methods, i.e., CSRec-M, CSRec-D, and CSRec-T, on Last.FM dataset with (a) different lengths, and (b) different levels of sparisty. We control the sparsity by randomly removing different proportions of users. The sparsity is 96.1\%, 96.7\%, 97.1\% and 97.3\% respectively. We choose GRU4Rec~\citep{gru4rec} as the base model.}
    \label{fig:abla-effect-len-and-sp}
\end{figure}

We also study how the sequence length of the user-item interactions affects the methods in Figure~\ref{fig:abla-effect-of-length}.
Users with more interaction histories will not necessarily get a more satisfied recommendation according to \citep{exp:effect-of-length}.
However, training with our proposed methods leads to increased resilience to the length of interaction history. 
This observation indicates that the proposed CSRec can consistently improve the recommendation performance given different lengths of the interaction sequences.

Sparsity is often a concern in the recommendation literature, and we conduct an experiment, in Figure~\ref{fig:different-sparsity}, to show that training with our methods is relatively robust to different levels of sparsity.
Training with CSRec-M outperforms the others settings across all the different levels of sparsity.
Even though do not preserve their performance at $\text{sparsity}=97.3\%$, our other two training methods, i.e., CSRec-D and CSRec-T, still show more promising results than the Base model and SoftRec on other sparsity settings.

\begin{table*}
        \caption{Comparison of the top-$n$ recommendation performance of different models ($n=10$) on two rating datasets using the filtered metrics described in Section~\ref{exp:filtered-metrics}. RC$^+$ is short for Recall$^+$@10. NG$^+$ is short for NDCG$^+$@10. \textbf{Boldface} denotes the highest score. $**$ and $*$ denote significant improvements over the corresponding SoftRec baseline ($p<0.05$ and $p<0.1$, respectively).}
        \centering
        \setlength{\columnsep}{0.25cm}
        \small
    
    \begin{tabular}{ll cc cc cc cc}
        \toprule    
         & & \multicolumn{2}{c}{GRU4Rec}&\multicolumn{2}{c}{NextItNet}&\multicolumn{2}{c}{NARM}&\multicolumn{2}{c}{SASRec}\\
        \cmidrule(lr){3-4}\cmidrule(lr){5-6}\cmidrule(lr){7-8}\cmidrule(lr){9-10}
         \bf Dataset & \bf Method &RC$^+$(\%) & NG$^+$(\%) &RC$^+$(\%) & NG$^+$(\%)&RC$^+$(\%) & NG$^+$(\%)&RC$^+$(\%) & NG$^+$(\%)\\
        \midrule
        \multirow{5}*{Electronics} & 
        Base & 5.577 & 3.074 & 5.111 & 3.344 & 5.575 & 3.262 & 7.394 & 4.279 \\
        & SoftRec & 5.812 & 3.267 & 5.484 & 3.608 & 5.902 & 3.435 & 7.434 & 4.359 \\
        &CSRec-M & 5.937\rlap{**} & 3.358\rlap{**} & \textbf{5.946}\rlap{**} & \textbf{3.968}\rlap{**} & 6.195\rlap{**} & 3.673\rlap{**} & 7.235 & 4.239 \\
        &CSRec-D & 5.901\rlap{*} & 3.319\rlap{**} & 5.555 & 3.665\rlap{**} & 6.250\rlap{**} & 3.720\rlap{**} & 7.207 & 4.209 \\
        &CSRec-T & \textbf{5.975}\rlap{**} & \textbf{3.381}\rlap{*} & 5.635\rlap{**} & 3.752\rlap{**} & \textbf{6.329}\rlap{**} & \textbf{3.769}\rlap{**} & \textbf{7.453} & \textbf{4.388}\\
        \midrule
        \multirow{5}{*}{\begin{tabular}{@{}l@{}}Movies and \\TV\end{tabular}} & 
        Base & 8.603 & 4.927 & 7.554 & 4.588 & 9.712 & 5.794 & 11.608 & 7.071 \\
        & SoftRec & 9.818 & 5.599 & 8.388 & 5.120 & 9.727 & 5.705 & 11.296 & 6.874 \\
        &CSRec-M & 9.925\rlap{**} & 5.743\rlap{**} & \textbf{8.992}\rlap{**} & \textbf{5.503}\rlap{**} & \textbf{10.361}\rlap{**} & \textbf{6.190}\rlap{**} & 11.641\rlap{**} & 7.076\rlap{**} \\
        &CSRec-D & 9.738\rlap{*} & 5.656\rlap{*} & 8.558 & 5.190 & 10.079\rlap{**} & 6.015\rlap{**} & \textbf{11.641}\rlap{*} & \textbf{7.087}\rlap{*} \\
        &CSRec-T & \textbf{9.958}\rlap{**} & \textbf{5.749}\rlap{**} & 8.643\rlap{**} & 5.265\rlap{**} & 10.175\rlap{**} & 6.051\rlap{**} & 11.604 & 7.045\\
       \bottomrule
        \end{tabular}
        \label{tab:robust_result}
    \end{table*}

To conclude, our proposed training methods not only lead to improved performance in standard evaluation scenarios but also to better outcomes on the positive user preference evaluation. The teacher module of CSRec can effectively help the student recommender generate recommendations that lead to real positive user preferences given corrupted training data and thus improve the robustness of the student module. Besides, the proposed CSRec also shows promising improvements in alleviating the popularity bias and provides robust performance in different settings of sequence lengths and sparsity levels.

\begin{figure*}
\centering
    \begin{minipage}{\linewidth}
        \captionsetup{labelformat=empty}
        \centering
        \begin{minipage}{0.9\linewidth}
            \captionsetup[subfigure]{skip=-1pt}
            \subcaptionbox{Electronics}{\includegraphics[width = 0.45\linewidth]{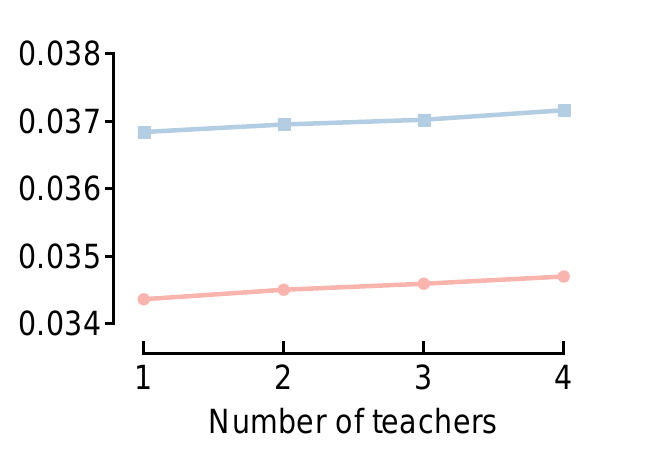}}
            \subcaptionbox{Movies and TV}{\includegraphics[width = 0.45\linewidth]{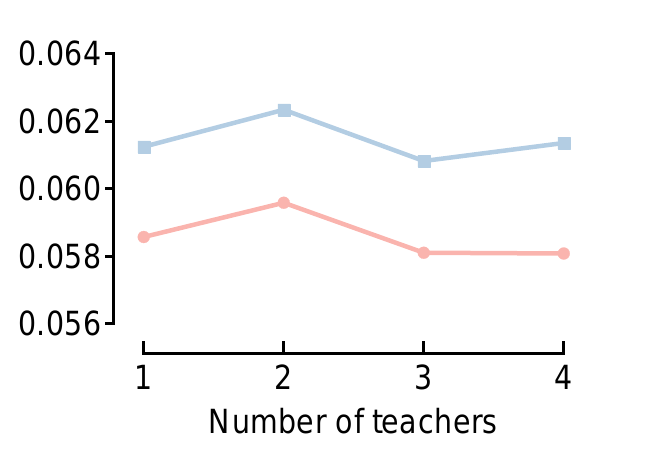}}
            \captionsetup{skip=3pt}
            \caption*{CSRec-M}
        \end{minipage}
        \begin{minipage}{0.9\linewidth}
            \captionsetup[subfigure]{skip=-1pt}
            \subcaptionbox{Electronics}{\includegraphics[width = 0.45\linewidth]{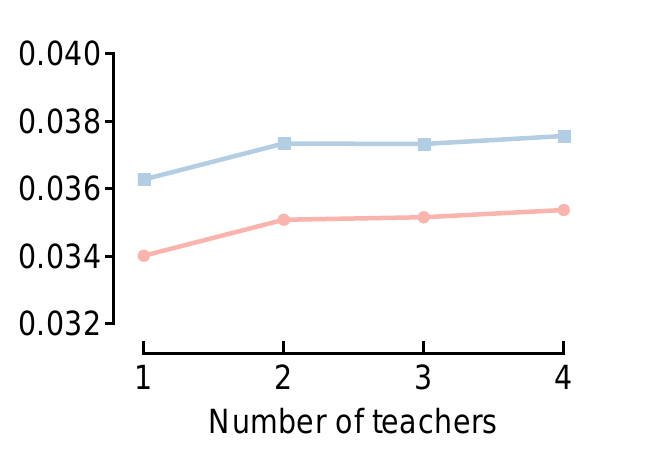}}
            \subcaptionbox{Movies and TV}{\includegraphics[width = 0.45\linewidth]{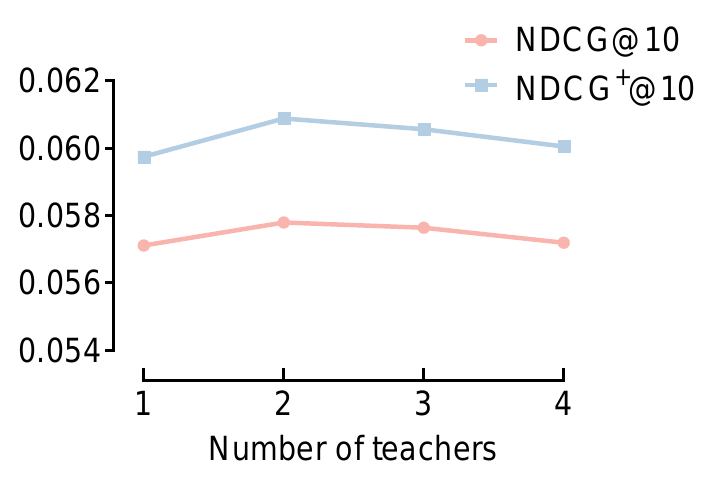}}
            \captionsetup{skip=3pt}
            \caption*{CSRec-D}
        \end{minipage}
    \end{minipage}
    
\captionsetup{skip=3pt}
\caption{Effect of the number of teacher models using NARM as the student model.}
\label{fig:abla-number-of-teacher}
\end{figure*}

\subsection{Ablation study}

In this section, we study how the design of the teacher module affects the performance of students.
Since both CSRec-M and CSRec-D use an ensemble of multiple teachers, we first look into the effect of the number of teachers (i.e., $m$). 
We report the results using NARM as the student model on two Amazon datasets, as shown in Figure~\ref{fig:abla-number-of-teacher}. Results for other student models show similar trends. From the results, we can see that the performance is
improved when the number of teachers is increased from 1 to 2, which means the student can benefit from multiple teachers. However, further increasing the number of teachers introduces little improvement, which means that more teachers are not necessarily giving us better soft labels and recommendation results.

\begin{figure}
    \centering
    \begin{minipage}{\linewidth}
    \centering
        \captionsetup{labelformat=empty}
        \begin{minipage}{\linewidth}
        \centering
            \captionsetup[subfigure]{skip=-4pt}
            \subcaptionbox{Electronics}{\includegraphics[width = 0.4\linewidth]{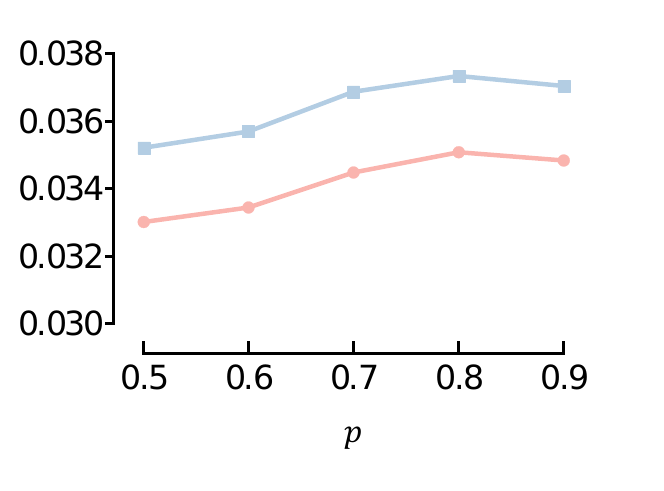}}
            \subcaptionbox{Movies and TV}{\includegraphics[width = 0.41\linewidth]{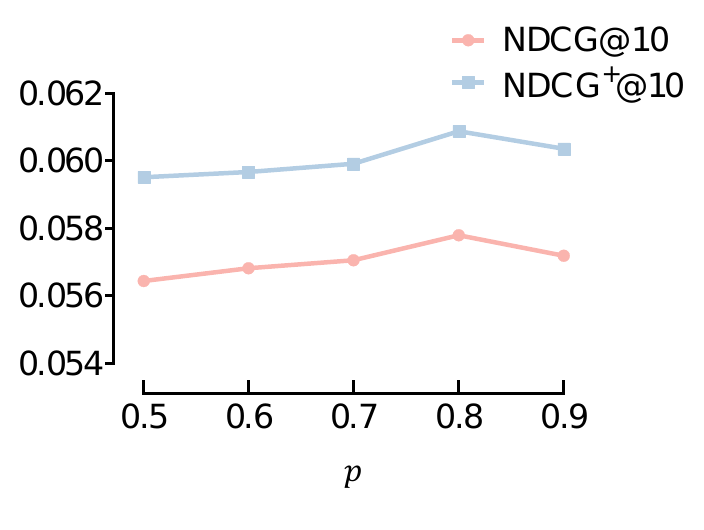}}
        \end{minipage}
    \end{minipage}
    \captionsetup{skip=4pt}
    \caption{Effect of the sub-sampling ratio $p$ in CSRec-D, using NARM as the student model.}
    \label{fig:abla-sampling-ratio}
\end{figure}

For CSRec-D, we also investigate the effect of the sub-sampling ratio. 
The result of using NARM as the student model is shown in Figure~\ref{fig:abla-sampling-ratio}. A large sampling ratio (i.e., $p>0.8$) will not always give us better students. Larger sub-datasets have more overlapping samples, making the teacher modules more similar to each other and sharing the same data-level bias and noise. 
However, a sampling ratio that is too small introduces too much randomness and also downgrades the performance. $p=0.8$ is a safe and robust choice.

\begin{figure}
    \centering
    \begin{minipage}{\linewidth}
        \captionsetup{labelformat=empty}
        \begin{minipage}{\linewidth}
        \centering
            \captionsetup[subfigure]{skip=0pt}
            \subcaptionbox{Electronics}{\includegraphics[width = 0.42\linewidth]{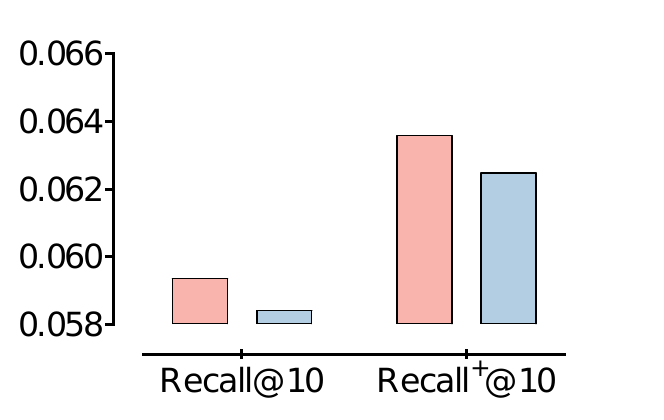}}
            \subcaptionbox{Movies and TV}{\includegraphics[width = 0.42\linewidth]{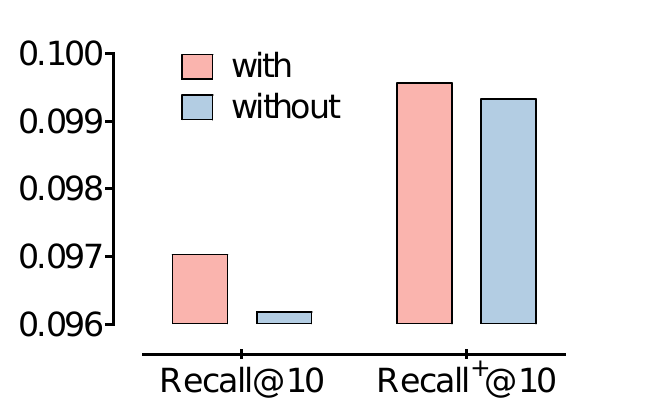}}
        \end{minipage}
    \end{minipage}
    \captionsetup{skip=4pt}
    \caption{Effect of the term $- \sum_{j\in \mathcal{I}}\log(h_{i,j}) P_{g_1}(j\mid\mathbf{s}_u)$ in the robust loss $\ell_r$ of CSRec-T. Here, ``with'' and ``without'' denote whether the term is used or not. The student model used is NARM.}
    \label{fig:abla-expectation-term-in-TL}
\end{figure}

Finally, for CSRec-T, we investigate the effect of the robust training loss $\ell_r$ (Eq.~\ref{eq:dvae_loss}).
Since $\ell_r$ is similar to the regular distillation loss except for the  term $- \sum_{j\in \mathcal{I}}\log(h_{i,j})P_{g_1}(j\mid \mathbf{s}_u)$, we thus check the effectiveness of this expectation term and report the results in Figure~\ref{fig:abla-expectation-term-in-TL}, using NARM as the student model.
We see that the expectation term significantly improves the student's performance, which demonstrates the effectiveness of our proposed methods.

\section{Conclusions}

We have proposed a new learning framework, CSRec, to train a robust sequential recommender from noisy, implicit feedback. The key idea is to introduce confident, soft labels to provide robust guidance in the learning process. 
We have presented three teacher modules, i.e., model-level, data-level, and training-level, for generating high-quality and confident, soft labels from noisy user-item interactions.
We have conducted extensive experiments to assess the effectiveness of the proposed learning framework. 

Experimental results on four datasets and diverse student models demonstrate that training with the proposed learning framework CSRec helps to improve the recommendation performance.
It can be applied to various deep-learning-based sequential recommendation models.

The broader implications of our work are that we have demonstrated the potential of soft labels for training sequential recommender systems, which we believe should further investigated and developed as a generic framework for producing better soft label guidance for a broader set of recommendation tasks.

Limitations of our work concern the inefficiency of the whole training framework pipeline due to the required multiple well-trained teacher models. 
The lack of exploration of different ensemble methods for further analysis and comparison is also one of the limitations.

As to future work, we see a number of promising directions. 
First, more intuitive and general teacher modules could be designed beyond the ones we considered in this paper, perhaps targeting specific types of noise or bias present in logged interaction data. 
Second, we aim to develop a more in-depth analysis to see how soft labels intrinsically affect the student model learning.

\begin{acks}
This research was funded by the Natural Science Foundation of China (62272274, 61972234, 62072279, 62102234, 62202271),
Meituan, the Natural Science Foundation of Shandong Province (ZR2022QF004), 
the Key Scientific and Technological Innovation Program of Shandong Province (2019JZZY010129), Shandong University multidisciplinary research and innovation team of young scholars (No.~2020QNQT017), 
the Tencent WeChat Rhino-Bird Focused Research Program (JR-WXG2021411), 
the Fundamental Research Funds of Shandong University, 
the Hybrid Intelligence Center, a 10-year program funded by the Dutch Ministry of Education, Culture and Science through the Netherlands Organization for Scientific Research, \url{https://hybrid-intelligence-centre.nl}, 
project LESSEN with project number NWA.1389.20.183 of the research program NWA ORC 2020/21, which is (partly) financed by the Dutch Research Council (NWO), 
and
the FINDHR (Fairness and Intersectional Non-Discrimination in Human Recommendation) project that received funding from the European Union’s Horizon Europe research and innovation program under grant agreement No 101070212.

All content represents the opinion of the authors, which is not necessarily shared or endorsed by their respective employers and/or sponsors.
\end{acks}

\bibliographystyle{ACM-Reference-Format}
\bibliography{references}

\end{document}